\documentclass[journal]{IEEEtran} 
\usepackage[T1]{fontenc}
\usepackage{amsmath,amsfonts}
\usepackage{algorithmic}
\usepackage{algorithm}
\usepackage{array}
\usepackage{color}
\usepackage[caption=false,font=normalsize,labelfont=sf,textfont=sf]{subfig}
\usepackage{textcomp}
\usepackage{stfloats}
\usepackage{url}
\usepackage{verbatim}
\usepackage{graphicx}
\graphicspath{{img/}}     
\usepackage{cite}
\usepackage{tikz}
\usepackage{hyperref}
\usepackage{tikz}
\usetikzlibrary{calc,trees,positioning,arrows,chains,shapes.geometric,%
	decorations.pathreplacing,decorations.pathmorphing,shapes,%
	matrix,shapes.symbols,shapes,arrows,calc,backgrounds,fit,decorations.pathreplacing, intersections}

\begin{document}

\title{Ensured Energy: How a Serious Game can Reach and Engage Diverse Societal Groups in Swiss Energy Transition}


\author{Toby Simpson, 
        Saara Jones,
        Gracia Br{\"u}ckmann, 
        Walid El-Ajou, 
        Erwan Moreira, 
        Borja Martinez Oltra,
        Rolf Krause, 
        Michael Multerer,
        Isabelle Stadelmann 
        \thanks{T.\ Simpson, S.\ Jones E.\ Moreira, B.\ Martinez Oltra 
          and M.\ Multerer are with the Universit{\`a}  della Svizzera italiana in Switzerland.} 
        \thanks{W.\ El-Ajou, G.\ Br{\"u}ckmann, I.\ Stadelmann are with the 
        University of Bern in Switzerland.}
        \thanks{R.\ Krause, is with the Universit{\`a} della Svizzera italiana and UniDistance Suisse in Switzerland and King Abdullah University of Science and Technology in Saudi Arabia} 
      }


      \markboth{Ensured Energy}{}


\maketitle
\begin{abstract} In support of Switzerland's energy and climate strategy for 2050, researchers investigate scenarios for the transition of energy systems towards a higher share of renewables, assessing their social, environmental and economic impact. Their results guide stakeholders and policy makers in designing resilient and sustainable systems. A crucial condition to successfully implement these solution in the real-world is that the population supports these transitions. Social scientists have identified the high complexity of energy systems and energy policy as one reason why popular support for the implementation of the energy transition is often limited. This paper proposes serious gaming as a novel approach to inform and sensitize a broader public for the transition of the energy system in Switzerland. We motivate and describe the design of an online game in which players experience an accurate simulation of current and future energy provision and manage transition towards a sustainable future. We present the embedding of this serious game into a large-scale population survey and report findings on player characteristics and engagement. We show that a serious game can successfully attract participants from diverse societal groups and highlight the challenge of balancing complexity and entertainment.
\end{abstract}


\section{Introduction}

\subsection{The role of serious gaming in opinion formation about the energy transition}


To reach the Swiss goal of a net-zero energy system compatible 
with the 2015 Paris Agreement\cite{sfoe_swiss_federal_office_of_energy_langfristige_2019,Baker2025}, different pathways\cite{Panos2023} towards energy transition are simulated and evaluated in terms of their potential effects on society and the environment. While one criterion is set, namely that these pathways should lead to a carbon-neutral
energy system by 2050, they differ in other important aspects\cite{trutnevyteRenewableEnergyOutlook2024}. For example, they emphasize varying energy technologies, rely on varying degrees of energy imports, or use different policy measures to generate incentives for an accelerated transition or for reduced energy consumption. 

While this research suggests that different paths to net-zero are both technically possible and feasible for Switzerland, one of the main challenges to implementation seems to be a lack of social acceptance \cite{Dermont2017BringingAcceptance}. This challenge is not specific to Switzerland. Indeed, public resistance to transition towards net-zero energy systems that mitigate climate change has become a polarized issue involving electoral risks for established parties and potential for democratic backlash \cite{Bosetti2025}. However, what is specific to the Swiss case is that the participatory context provides citizens with ample possibilities to directly decide on relevant legislation and block renewable energy projects \cite{Stadelmann-Steffen2011}. 

To better anticipate and plan the implementation of renewable energy pathways, it is therefore crucial to better understand and include citizens' perceptions with regard to future energy systems\cite{gemma_delafield_conceptual_2021, PERLAVICIUTE2014361,
yann_blumer_two-level_2018}. One finding of existing research is that people are generally in favour of a carbon neutral energy system \cite{trutnevyteRenewableEnergyOutlook2024}, but are reluctant to support projects and policies aimed
at transition towards renewables
\cite{Stadelmann-Steffen2018a}. This apparent mismatch can be attributed in part
to the high complexity of the topic, making it difficult for people to
understand and recognize the benefits of net zero policies and technologies
\cite{Stadelmann-Steffen2018,Stoutenborough2014TheComparison}. Even more so, the evaluation of \textit{transition pathways} (not single policy measures) towards net-zero involves questions around complex concepts over long time-frames, and includes a potentially infinite number of different hypothetical states. 
At the same time, research
shows that simple information provision on these complex issues is often
insufficient to effectively engage key social groups or to build greater support
for transition measures \cite{Buergisser2024,Fremstad2022}. 

\begin{figure}[t]
    \centering
    \includegraphics[width=\linewidth]{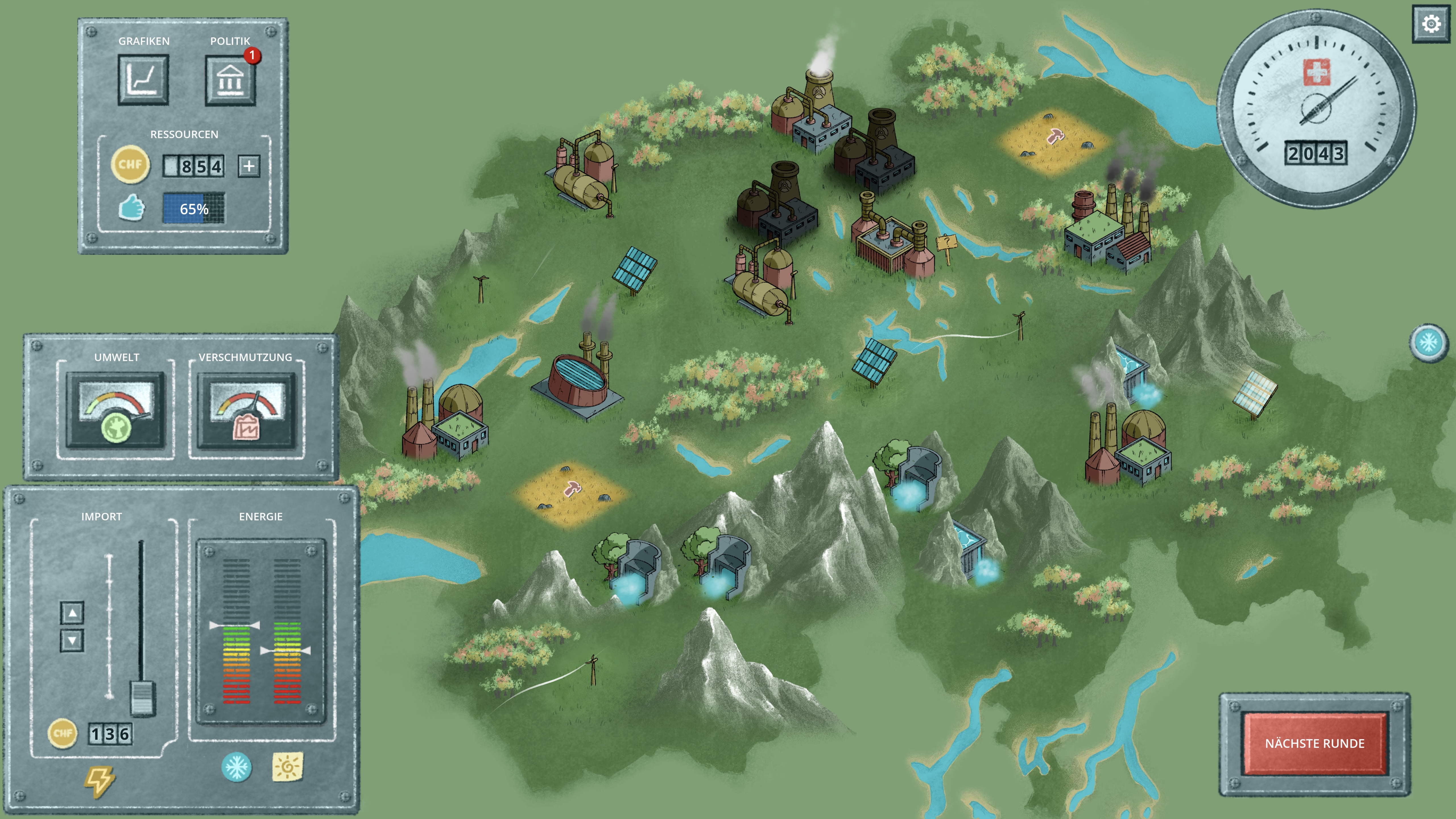}
    \caption{Screenshot from \textit{Ensured Energy}.}
    \label{fig:screenshot1}
\end{figure}

This is the starting point of the present study, which proposes serious gaming as a tool to sensitize the broader public to energy transition. Serious games allow players to immerse themselves in scenarios and are thus beneficial in gaining attention, simplifying information processing, and
inducing emotion towards decision-making. This approach is expected both to facilitate opinion formation in players and to contribute to a more accurate analysis of preferences \cite{aubert_review_2018,
kahneman2003maps, Chapman_Johnson_2002,
doi:10.1146/annurev-psych-010213-115043}. Furthermore, through the use of a
serious game, a variety of scenarios and disruptive events can be presented, testing how preferences for the configuration of an energy system are context dependent \cite{gordon_serious_2020, yiannakoulias_decision_2020}. 

One crucial precondition for serious games to unfold these positive properties is that they are able to attract and engage players. 
Therefore, the guiding question of the present study is: \textit{Can a serious game about energy transition function as an effective tool to engage a wider audience?} Investigating this question provides insights on the potential of serious games to reach the broader population. Only if players really engage with the game and if game engagement goes beyond the typical gamers' group of young male individuals, we can expect that a serious game could eventually ``make a difference'', e.g., by enhancing the population's awareness and comprehension of a complex issue like energy transition.

To answer the question, we present the development and design of the online game \textit{Ensured Energy} (see Figure \ref{fig:screenshot1}) in which players experience an accurate simulation of current and future energy provision and are asked to secure Switzerland's energy supply until 2050, thereby managing the transition towards a sustainable future. In addition, we propose a novel approach to embed the serious game into a large-scale population survey as a randomized intervention, which enables us to test in how far it is able to engage the broader population. More specifically, we use the information collected in the survey to study characteristics of the players compared to non-players. Furthermore, we can compare the treatment group (who was sent to the game in the middle of the survey) and the control (who was only informally invited to the game at the end of the survey) with respect to their game engagement.


Our approach extends previous methods in public opinion research in two distinct ways. Firstly, the game can be conceptualized as a sophisticated information treatment, providing participants with structured insights into complex energy transition pathways. Our study informs about the potential of such games to indeed ``treat'' respondents of a broader population sample. Secondly, the game itself can capture context-specific preferences by immersing participants in hypothetical but realistic future scenarios. 

\subsection{Serious games as sophisticated information treatment}


The complex topic of energy transition reduces the ability of survey respondents
to grasp the pros and cons of various measures and the trade-offs involved in
the decision-making process \cite{Stadelmann-Steffen2018,
Stoutenborough2014TheComparison}. Serious games can mitigate this issue and
are widely used to foster learning \cite{Keusch2017, Goldstein2013}. 
In general, the potential to make complex topics more understandable is seen as an advantage of serious games
\cite{bekebrede_towards_2018, aubert_review_2018, ampatzidou_mapping_2019}. Player
engagement can change perceptions \cite{Kazhamiakin2015,Schiele2018} and preferences \cite{Cechanowicz2013,Sailer2017}, but it can improve efficacy, which is defined as ``beliefs in one's
capabilities to execute the competencies needed to exercise control over events
that affect one's welfare'' \cite{Bandura1986}. Serious games that provide feedback have the potential to allow survey respondents to make informed decisions even in situations of high complexity \cite{Harms2015,Aubert2018, abad_assessing_2020}. 

In the context of \textit{Ensured Energy}, players manage the Swiss energy system, with the aim of ensuring the energy supply for Switzerland until 2050. For this purpose, they can, for example, construct new power plants, take political measures or decide to import more energy from neighbouring countries. During the game, players are shown indicators relating to the energy system and its economic and environmental impact. These indicators effectively confront the player with a trade-off, since it is not possible to choose a single best option. All
potential decisions come with advantages and disadvantages at the level of
certain indicators. Thus, we can investigate how different contexts, provided
by the game and mimicking realistic real-world situations, influence the
decisions of the respondents.  

\subsection{Outline}

The rest of the paper is organised as follows: Section \ref{sec:game} describes
the design of the game and the gameplay itself in more detail. Section \ref{sec:calc} presents the calculation model that underlies the game and its calibration to official statistics. Section \ref{sec:data} includes details of how the game was integrated into a public opinion survey and presents empirical results to answer our research question. Finally, Section \ref{sec:conc} gives an overall assessment of the game and an
outlook for applications and future work.

\section{Gameplay Design}\label{sec:game}

\subsection{Gameplay Objectives}

We adhere to the Gameplay/Purpose/Scope (GPS) framework, 
which states that serious games have three different purposes: to communicate, to educate, and to collect data \cite{Djaouti2011}. \textit{Ensured Energy} informs players about the potentials and
limitations of the Swiss energy system and collects their priorities
regarding energy transition through their in-game decisions. The latter is the
most relevant aspect of the game, effectively performing a Multiple-Criteria
Decision Analysis (MCDA) \cite{RePEc:wsi:wsbook:8042} study with lay subjects. Although the game is not designed as an educational tool in the narrow sense, it contains informative elements and feedback relating to the Swiss energy landscape.

When it comes to data collection, agency and the ability to choose between
different options are crucial in understanding preferences. The fictive point of view of a national energy manager who has to secure energy supply for Switzerland is therefore chosen, allowing players to plan energy transition, manage different resources, and trade-off
outcomes. The gameplay of simulation/management games fits this frame 
\cite{Djaouti2011}, although some of the main characteristics of the genre had to be abandoned in order to give a more realistic context to the game. The traditional Tabula Rasa or blank slate of simulation games \cite{Molleindustria2017} did not
make sense in a game that replicates the current energy system of Switzerland. The game therefore is initialised with a Swiss map that contains multiple power plants representing the current reality. Players allocate capacity for power generation to nuclear, hydroelectric, solar, wind, biomass, biogas and waste.
 Even though the
Swiss map is used as the background of the game interface, the geographical
location of these sites has no impact on the constructed power plants, both as a
simplification and to avoid regional bias. Another necessary change was in resource
management. While there has been a shift in the goals and game loops of
simulation games recently, the most common purpose is still growth (population,
economic, etc.). Growth in our game is not under the control of the player, instead their objective is balance, in the supply and demand of energy as well as in their financial budget.


Several game elements help shape a more realistic political landscape. 
Switzerland has decided to opt out of nuclear power \cite{Energiegesetz2017}, so there is no possibility
to build new nuclear reactors in the game. In addition, instead of unlimited unused space, as, e.g., in \textit{SimCity} \cite{simcity},
construction sites are limited, which integrates the important and strongly politicised trade-off between energy infrastructure and land use. Moreover, differing construction times for power plants are accounted for by type, with an appropriate delay between the player's decision to build and the moment at which a plant contributes to energy supply\cite{schmid2024}. 
Similarly, resistance to large solar parks requires the player to implement policies that allow an increased number of upgrades to solar plants in-game.

\subsection{Gameplay Overview}

The game is structured into 10 turns over 30 years and the player must provide sufficient energy for both summer and winter at each turn in order to move on to the next. 
The player has different options to reach the required energy demand: 
\begin{itemize}
    \item Select a free building area and build a new power plant.
    \item Increase or decrease the capacity of existing power plants.
    \item Import energy from abroad.
    \item Implement political measures (policies) that reduce demand or help build
          power plants more quickly.
    \item Hold political campaigns to increase the chances of policies being
          accepted.
\end{itemize}

Each power plant has characteristics such as financial cost per energy output,
construction time and impacts on different factors such as land use, emissions
and the seasonal availability of electricity. Policies have a probability of
acceptance, which can increase through information campaigns or be lowered
according to the player's choices for energy supply. Throughout the game, the
player must react to various random shocks, see Table~\ref{tab:shock1}, that
impact energy supply and demand, testing the resilience of their strategy. 

In the introductory text at the start of the game, the player is assigned one of two different frames: one instructs them solely to maintaining energy supply, while the other emphasises the need to also effectively manage the transition to renewables. Regardless of this context, all players are free to set their own goals, as long as there is sufficient energy supplied in each round. Each player is scored according to a set of metrics that are calculated and displayed at the end of the game. Scores are ranked and stored on a leader board that can be ordered by each of the metrics. The game is available in four languages: German, French, Italian and English.

\subsection{Game Implementation}

\begin{figure}[t]
\noindent
\begin{tikzpicture}[auto, box/.style = {draw,  thick, 
  rounded corners=5pt,font=\sffamily}]
\begin{scope}[xshift=-1cm]
    \node[anchor=south west, inner sep=0, opacity=0.75] (image) at (0,0) 
      {\includegraphics[width=\linewidth]{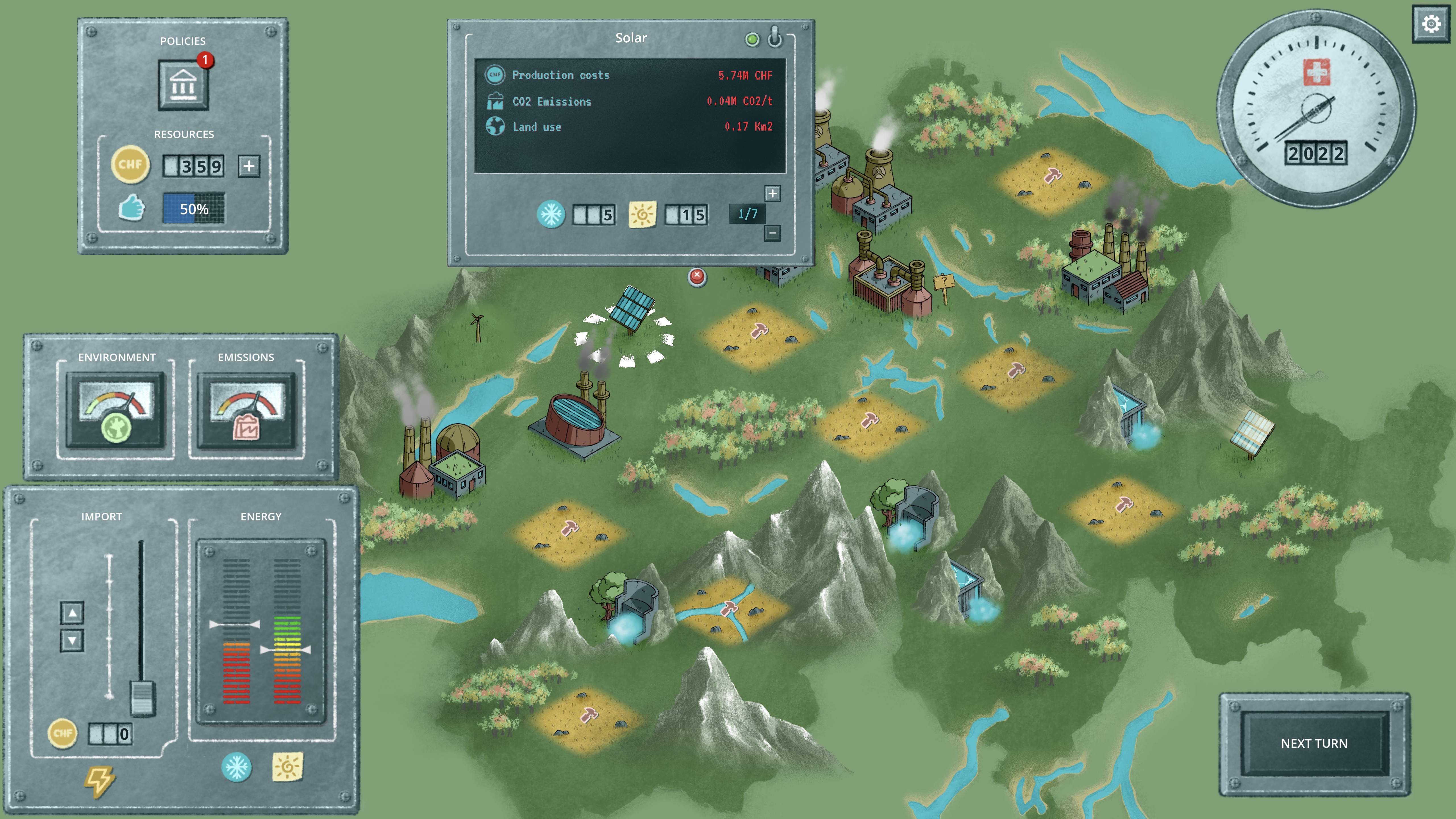}};
    \begin{scope}[x={(image.south east)},y={(image.north west)}]
	\draw[box] (0.00, 0.00) rectangle (0.25, 0.40)%
    node [anchor=center,midway] {A};
	\draw[box] (0.00, 0.41) rectangle (0.25, 0.60)%
    node [anchor=center,midway] {B};
	\draw[box] (0.05, 0.69) rectangle (0.20, 0.765)%
    node [anchor=center,midway] {C};
	\draw[box] (0.05, 0.775) rectangle (0.20, 0.85)%
    node [anchor=center,midway] {D};
	\draw[box] (0.05, 0.86) rectangle (0.20, 0.97)%
    node [anchor=center,midway] {E};
	\draw[box] (0.3, 0.8) rectangle (0.57, 0.98)%
    node [anchor=center,midway] {F};
    \draw[box] (0.3, 0.69) rectangle (0.57, 0.78)%
      node [anchor=center,midway] {G};
	\draw[box] (0.39, 0.53) rectangle (0.47, 0.66)%
    node [anchor=center,midway] {H};
	\draw[box] (0.45, 0.19) rectangle (0.55, 0.32)%
    node [anchor=center,midway] {I};
	\draw[box] (0.72, 0.31) rectangle (0.82, 0.44)%
    node [anchor=center,midway] {J};
	\draw[box] (0.83, 0.74) rectangle (0.98, 0.99)%
    node [anchor=center,midway] {K};
    \end{scope}
\end{scope}
\end{tikzpicture}%

\begin{center}
    \small
    \def\arraystretch{1.25}
    \begin{tabular}{ c|l } 
     \textbf{Key} & \textbf{Description}  \\ 
     \hline
    A & Import slider, summer/winter energy levels  \\ 
    B & Land use and emissions gauges  \\ 
    C & Political support bar  \\ 
    D & Budget and borrowing buttons \\ 
    E & Policy button and link to metric charts \\ 
    F & Cost and impact of selected power plant \\ 
    G & Capacity for summer/winter and upgrade button \\ 
    H & Selected power plant \\ 
    I & Water construction site \\ 
    J & Land construction site \\ 
    K & Current turn and year \\ 

    \end{tabular}
    \end{center}

\caption{Components of the game user interface}
\label{fig:user-interface1}
\end{figure}

\textit{Ensured Energy} was created using the Godot 4 game
engine \cite{godot} and deployed via a web browser to facilitate
navigation to and from the survey. During the development of the game, limitations around software led to a decision that the game should be developed for a desktop-only browser environment. This in turn guided the design of the user interface and gameplay, allowing a more detailed display and controls, but with unexpected consequences that are discussed in Section \ref{sec:insight1}.

The layout and function of the game controls
are shown in Figure \ref{fig:user-interface1}. Every player starts with the same scenario, which is the state of the Swiss energy system in 2022. For simplicity, the map shows one or two power plants of each type, with the sum of their capacities reflecting the current energy mix.

At each turn, the player is given a budget that can be used to
build new power plants, upgrade existing structures and import energy. Each
plant has a building cost, an upgrade cost and production cost, which is
taken automatically at every turn. The player can borrow money up to a fixed amount
per turn with the loan repaid automatically with interest at the subsequent turn.
All power plants have an impact on land use and emissions. The player can find
the relevant information by clicking on a specific station or by clicking on the
overall corresponding gauge for the total.

\begin{table}[htb]
    \begin{center}
    \small
    \def\arraystretch{1.25}
    \begin{tabular}{ l|r|r|c } 
     \textbf{Power Plant} & \textbf{Build} & \textbf{Upgrade} & \textbf{Total} \\ 
     \hline
     River      & 50\%  & 1\% & 10 \\ 
     Reservoir  & 50\%  & 1\% & 10 \\ 
     Solar      & 25\%  & 5\% & 7 \\ 
     Wind       & 100\% & 50\% & 7 \\ 
     Gas        & 100\% & 50\% & 0 \\ 
     Biomass    & 100\% & 50\% & 0 \\ 
     Biogas     & 100\% & 50\% & 0 \\
     Waste      & 25\%  & 6.25\% & 0 \\ 
    \end{tabular}
    \end{center}
    \caption{Build and upgrade parameters for power plants showing percentage
    increases to capacity and maximum number of upgrades. }
    \label{tab:build1}
\end{table}

Table \ref{tab:build1} shows percentage increase in initial
capacity that results from building or upgrading the various power 
installations, as well as the total number of upgrades that can be applied.
Most power plants are built immediately, but wind, river and reservoir
installations take 2, 3 and 6 turns respectively. 

Build times can be reduced by implementing policies which provide education or support for corresponding energy technologies. Players may try to implement one policy per turn. The list of policies, their effects, and their initial acceptance probability are summarised in Table \ref{tab:policy1}:

\begin{table}[htb]
    \begin{center}
    \small
    \def\arraystretch{1.25}
    \begin{tabular}{ l|l|c } 
     \textbf{Policy} & \textbf{Effect} & \textbf{Accept} \\ 
     \hline
     Enable alpine PV       & increase solar upgrades   & 80\%  \\ 
     Fast track wind parks  & lower wind build time     & 60\%  \\ 
     Wind park regulation   & increase wind upgrades    & 50\%  \\ 
     Building insulation    & lower household demand    & 70\%  \\ 
     Industry subsidy       & lower industry demand     & 60\%  
    \end{tabular}
    \end{center}
    \caption{Summary of policies showing description, effects and probability
    of acceptance.}
    \label{tab:policy1}
\end{table}

Decisions made in response to shocks affect the probabilities for future policy acceptance, with unpopular responses reducing future acceptance and vice-versa. Shocks are events that occur randomly during the game and that have an effect on supply, demand or political support. If the energy supply of the player can
withstand the shock, the player is rewarded with an increased budget and increased political support. The different types of shock are summarised in Table \ref{tab:shock1}.

\begin{table}[htb]
    \begin{center}
    \small
    \def\arraystretch{1.25}
    \begin{tabular}{ l|l } 
     \textbf{Shock} & \textbf{Effect} \\ 
     \hline
    Cold spell  & increase winter demand by 5\%   \\ 
    Heat wave  & increase summer demand by 5\%   \\ 
    Mass immigration  & increase demand by 5\%   \\ 
    Renewable support  & increase support by 10\%   \\ 
    Glaciers melting  & for user information only    \\ 
    Nuclear re-introduction  & gather user preference only   
    \end{tabular}
    \end{center}
    \caption{Summary of shocks showing description and effect.}
    \label{tab:shock1}
\end{table}


In the case of a cold spell or heat wave, the player is given a choice which
allows them to trade-off the effects with respect to political support,
emissions or financial resources. The choice to re-introduce nuclear power
always takes place on turn 5.

In addition to allocating resources to electricity generation, a player may
reduce emissions by carbon sequestration. In this case, players allocate the
capacity of carbon capture and storage at a price set in the model.

\section{Model and Calculation}\label{sec:calc}

\subsection{Energy Balance}

\begin{figure}[t]
    \centering
    \includegraphics[width=\linewidth]{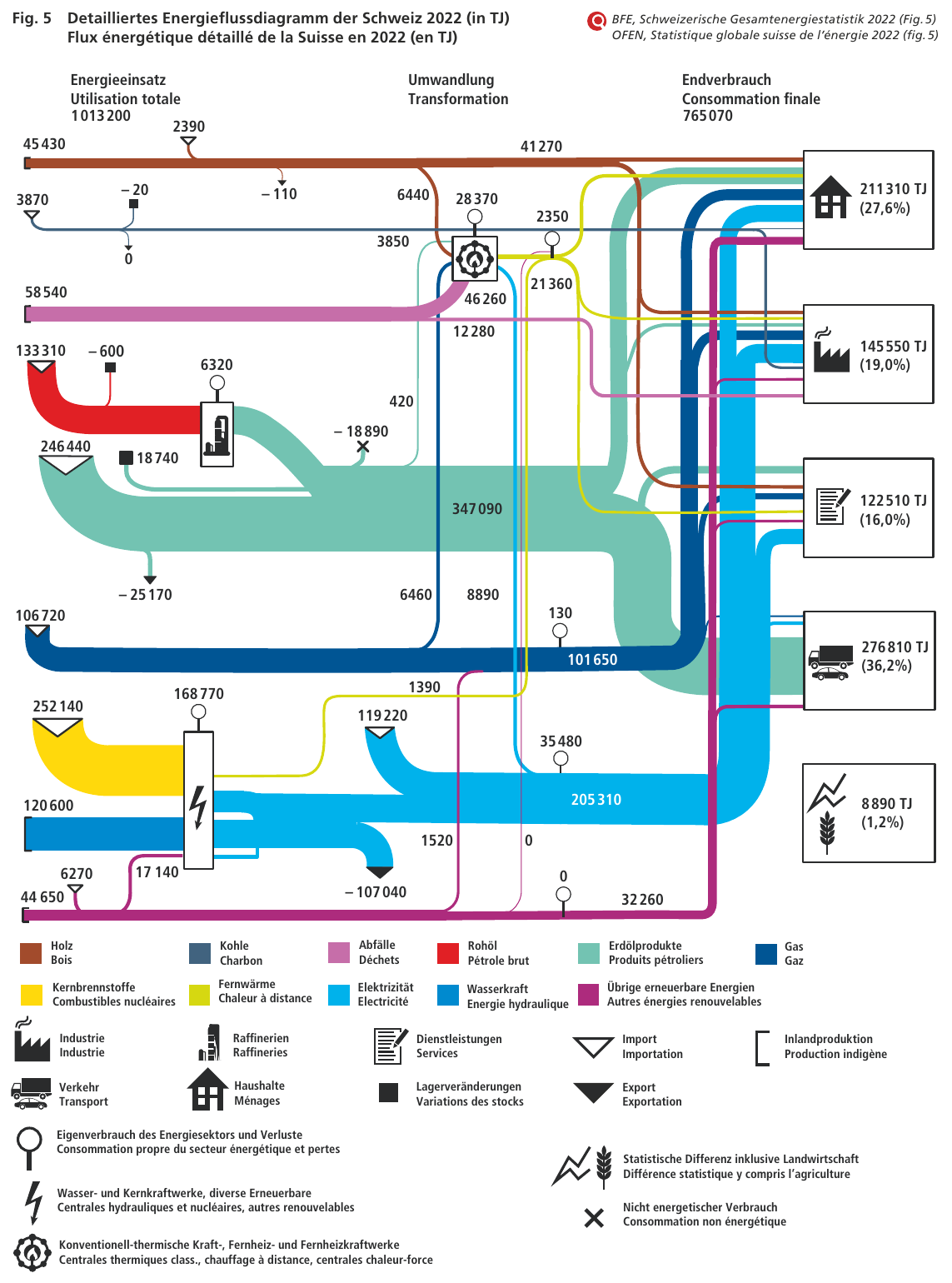} 
    \caption{Excerpt from \cite{bfe1} showing a schematic representation of 
    the energy balance for 2022.}
    \label{fig:bal1}
\end{figure}

To correctly record decisions, realistic questions must be posed. We use a 
reduced representation of the Swiss energy system that remains accurate with respect to the quantitative nature of the full model, while providing simplified and timely interaction for the player. The economic and environmental model that underlies the game takes its
structure from the annual reporting of the Swiss federal offices for energy\cite{bfe1}\cite{bfe2} and the environment\cite{bafu1}\cite{bafu2}. The energy balance is part of official statistics for most countries
\cite{Weber2022} and accounts for the flow of energy through the economy. A
schematic view is shown in Figure \ref{fig:bal1} and a brief description
follows: 

\begin{enumerate}
\item Primary energy enters the economy either as fuels
(oil, gas, nuclear) which may be imported, exported, produced domestically,
moved to and from stocks, or drawn directly from the environment (such as
hydroelectric, wind or solar energy).
\item Transformation converts primary energy in its raw state into other
forms, most importantly electricity. Installations such as nuclear or gas
fuelled power plants, hydroelectric facilities, solar panels or wind turbines
perform these functions. Conversion processes are not completely efficient,
and thus total energy is reduced during transformation.
\item Delivery of energy in its various forms to final users then takes
place, either via power distribution networks, pipelines, or by transport
of physical fuel. Delivery also incurs losses which must be
accounted for. 
\item Final consumption of energy, either in its raw or
transformed state is carried out by the five main economic sectors: households,
industry, services, transport and agriculture. Excess energy may be stored
or exported, and (most significantly for electricity) any shortfall must
be imported directly.
\end{enumerate}

\begin{figure*}
    \centering
    \includegraphics[width=0.8\linewidth]{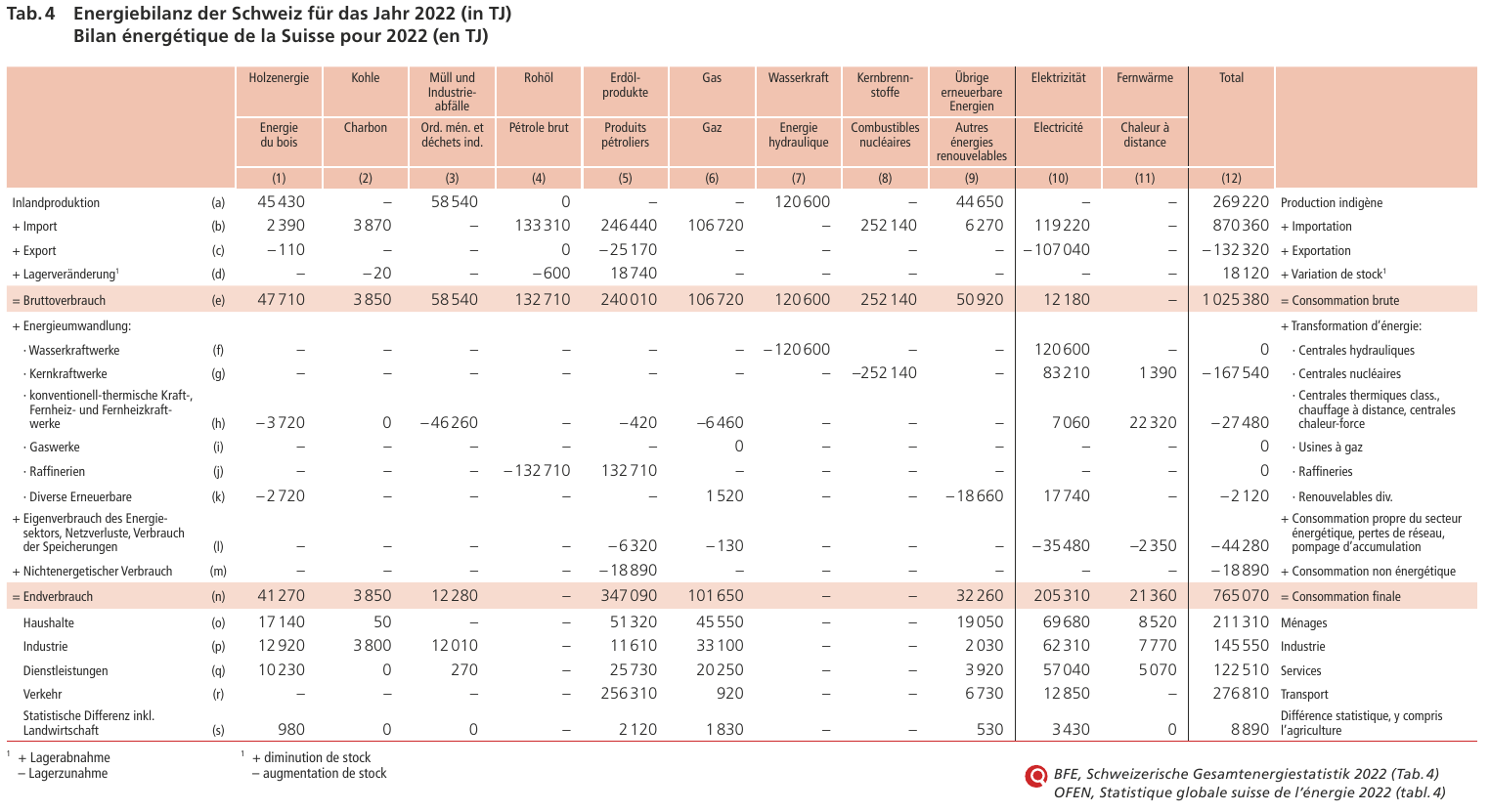}
    \caption{Excerpt from \cite{bfe1} showing a tabular representation of the
    energy balance for 2022.}
    \label{fig:bal2}
\end{figure*}

Figure \ref{fig:bal2} shows the same energy balance in tabular form \cite{bfe1},
where units are terajoules (1TJ = $10^{12}$ joules, 3.6TJ = 1GWh). Columns
represent energy in different forms, more detailed data for renewables were
obtained from the Swiss Federal Office of Energy (SFOE). Rows at the top of the
table represent quantities of primary energy available to the system. The central part of the table shows transformation processes moving quantities of energy from one column to another as they are converted. Losses due to delivery and processing are then shown, and in the lower part of the table are the quantities removed from the system by consumption in the five economic sectors.

Thus, all of the energy entering, leaving and transformed by the system is
accounted for. Changing these quantities over time while maintaining the balance
of energy allows the model to describe different scenarios for energy transition.

\subsection{Scenario Construction}
The model constructs a complete energy balance for each year of the game. Players install or remove power plants representing different energy conversion technologies. These changes in capacity affect the supply of final energy and demand for primary energy. Similarly, players may alter demand for final energy by enacting policies that subsidise renewable power generation or reduce consumption. Changes to supply and demand may also arise from shocks to the system as described earlier. The calculation of a synthetic energy balance thus takes place as follows:
\begin{enumerate}
    \item Given forecasts for final consumption, the total delivery requirement
      for each type of energy is calculated.
    \item Losses associated with storage and delivery of each energy type are
      added to give the amounts required after transformation.
    \item The quantities of each energy type made available by the various
      transformation processes are deducted from the amounts required.
    \item Any remaining unfulfilled requirement is satisfied by imports 
    and excesses of supply removed via exports.
\end{enumerate}
The approach reflects and extends the work of
\cite{Berntsen2017,Trutnevyte2016}, providing a complete description of the
Swiss energy balance over time.

\subsection{Calibration \& Forecasting}
\label{sec:calib1}

In order to reconstruct the energy balance at a future point in time, we require
estimates for levels of production, consumption, energy transformation, 
and the efficiencies and losses associated with each process. The model is calibrated to historic energy balance data
\cite{bfe1, bfe2} which includes time series for traditional sources since 1980
and renewables since 1990.

Since 2010 total final energy consumption in Switzerland has been decreasing on a linear trend, due mainly to increasing environmental temperatures\cite{bfe1}. In order to reproduce this overall trend, demand is projected using a linear regression over 8 years of time series data with supply and efficiency values held constant at their current time-averaged levels. The regression removes some of the effects of the COVID19 pandemic, which caused large distortions to recent data points dues to decreased economic activity. It also has the effect of damping the rapid growth in new technologies (e.g. biofuels, geothermal) which would otherwise be over-represented in future energy supply. 

The linear model therefore represents a compromise in design which favours slower change. Fewer parameters reduce uncertainty in estimation and computational complexity. The forecast thus provides a linear baseline to which shocks are then applied, allowing players to experience nonlinear changes in demand as events within the game. The problem for the player is thus not only to satisfy demand but to re-allocate resources to more sustainable and resilient forms of supply.



Historic data is annualised but in reality there are fluctuations in supply, demand and processing. Demand
for heating is greater in winter and cooling in summer. Solar and run-of-river have increased
availability in summer and nuclear power generation is therefore decreased during this period. Seasonal calibration uses hourly energy market data \cite{ec1} over a four year history. For annual seasonality the demand data was aggregated by quarter
and allocated to summer or winter but daily, weekly and intraday variations\cite{Weber2022} are beyond the
scope of the model.

\subsection{Metrics}

To assess and compare instances of the model and game, we calculate a set of
economic and environmental metrics which are shown to the players as final scores. Some metrics are intrinsic to the energy
balance itself, specifically net imports of fuel and electricity itself. These measure self-sufficiency in the energy economy and are a proxy measure of cost. We do not attempt to forecast energy market
prices but estimates of the cost of electricity generation (in CHF/TJ) are available in the literature
\cite{Xexakis2020,Ghodsvali2022,Holzer2023}. 

Emissions of greenhouse gases (CO\textsubscript{2}, N\textsubscript{2}O,
CH\textsubscript{4}, and SF\textsubscript{6}) related to both electricity
generation and consumption are calculated as an equivalent mass of
CO\textsubscript{2} in kg \cite{bafu1,bafu2,bafu3}. We also calculate equivalent
CO\textsubscript{2}-emissions for final consumption by combustion (fossil fuels,
wood, biofuels, biogas and waste). The values are not directly available but can
be calibrated from publicly available data \cite{bafu1,bafu2} using regression
as before.

Finally, we calculate a land use metric for energy transformation. This gives an
environmental opportunity cost associated with each type of electricity
generation. Values in km\textsuperscript{2}/TJ are available in the literature
\cite{Trutnevyte2016,Xexakis2020,Ghodsvali2022,Holzer2023}.

The metrics are calculated annually and summed over the period of the
simulation. They are summarised in Table \ref{tab:metric1}.

\begin{table}[htb]
    \begin{center}
    \small
    \def\arraystretch{1.25}
    \begin{tabular}{ l|l } 
     \textbf{Metric} & \textbf{Units}  \\ 
     \hline
    Nuclear fuel consumption                    & TJ   \\ 
    Fossil fuel consumption                     & TJ   \\ 
    Electricity imports                         & TJ   \\ 
    Emissions (CO\textsubscript{2} equivalent)  & t (mln)   \\ 
    Investment costs                            & CHF (mln)   \\ 
    Land use                                    & km\textsuperscript{2}   \\ 
    Seasonality (summer/total)                  &  \%
    \end{tabular}
    \end{center}
    \caption{Summary of metrics calculated per simulation with units.}
    \label{tab:metric1}
\end{table}

\subsection{Model Implementation}

The model is implemented on a web server using the LAMP architecture.
Interaction is via HTML, XMLHTTP or JSON interfaces. Details of the API can be
provided on request. Actions made by players are stored in an SQL database along
with cookies that allow cross-referencing of game instances with survey data.

\section{Integration of Survey and Game}
\label{sec:data}
\subsection{Survey Overview}
In order to evaluate whether the game can effectively engage players and enhance their awareness and comprehension of energy transition, in accordance with our research question, the game was embedded in a large-n survey study and used as an experimental treatment. The sample for the project was drawn from the
official Swiss population register. The Swiss Federal Statistical Office provided the contact addresses of 6,000 randomly selected Swiss residents
between the ages of 18 and 75. We used a stratified sample across the 26 Swiss cantons to account for differences in population size and to ensure
representation from all cantons. Respondents were recruited through a postal invitation containing a link to an online survey and a personal identification
token. The self-administered survey was prepared in German, French, Italian, and English using \textit{Qualtrics} software, with essential infrastructure and support provided by the Decision Sciences Laboratory\footnote{\url{https://www.descil.ethz.ch/}} at ETH Zurich.
All respondents entered the survey and filled out a first survey block assessing pre-treatment variables (e.g., environmental attitudes, perceived efficacy of individuals and the society to deal with the energy transition). After this first survey block respondents were assigned either to treatment or control group. For the assignment, a blocked randomization approach was based on respondents age, gender, language and pre-treatment individual and collective efficacy levels. Blocking ensures that respondents are precisely randomized on the included variables, while maintaining average randomization on all other variables
\cite{King_et_al_avoiding_2011}. The treatment group was then invited to play
the game before continuing with the second part of the survey. The control group directly continued with the second part of the survey. Only after finishing the survey, the control group was invited to play the game.

Using a personal identification token, we combined survey responses with information on decisions taken by respondents during the game. Specifically, the game recorded parameter
changes, including adjustments in energy production for different technologies
and changes in various indicators at each round. In addition, the game stored
information about shocks and policies, including their timing, options provided to players, and the responses they chose.

Respondents’ survey answers, including socio-demographic characteristics and attitudes, can thus be linked to their in-game decisions. This enables us to explore and compare how different groups interact with and respond to the game environment. 

\subsection{Participant Flow and Attrition}

Figure~\ref{fig:attrition} illustrates the study flow and attrition across the different stages of participation. Of the 6,000 individuals contacted through the Swiss population register, 2,164 accessed the survey and 2,029 completed the first survey block. A total of 2,024 respondents were successfully randomised into the treatment or control group. Among these, 1,453 launched the game and 542 completed at least one round. Overall, 354 respondents reached the final game round, and 1,758 respondents across both groups completed the second part of the survey (1,007 in the control group and 751 in the treatment group). Attrition occurred primarily between completing the first survey block and launching the game, as well as between launching the game and completing the first round. Because the survey and the game were self-administered and voluntary, detailed reasons for dropout are not systematically available.

\begin{figure}[ht]
\centering
\resizebox{0.9\columnwidth}{!}{%
\begin{tikzpicture}[
  box/.style={draw, rounded corners, align=center, minimum width=4cm, minimum height=.8cm},
  arrow/.style={->, thick}
]

\node[box] (sample) {Sampled from population\\(n = 6,000)};
\node[box, below=of sample] (access) {Accessed survey\\(n = 2,164)};
\node[box, below=of access] (block1) {Completed survey block 1\\(n = 2,029)};

\node[box, below left=1cm and 2cm of block1] (control) {CONTROL group};
\node[box, below right=1cm and 2cm of block1] (treat) {TREATMENT group};

\node[box, below=of control] (c2) {Completed survey block 2\\(n = 1,007)};
\node[box, below=of c2] (cgame) {Launched game\\(n = 706)};
\node[box, below=of cgame] (cround1) {Completed $\geq 1$ round\\(n = 139)};
\node[box, below=of cround1] (ccompleted) {Completed game\\(n = 101)};

\node[box, below=of treat] (tgame) {Launched game\\(n = 747)};
\node[box, below=of tgame] (tround1) {Completed $\geq 1$ round\\(n = 403)};
\node[box, below=of tround1] (tcompleted) {Completed game\\(n = 101)};
\node[box, below=of tcompleted] (t2) {Completed survey block 2\\(n = 751)};

\draw[arrow] (sample) -- (access);

\draw[arrow] (access) -- (block1);
\draw[arrow] (block1) -- (control);
\draw[arrow] (block1) -- (treat);
\draw[arrow] (control) -- (c2);
\draw[arrow] (c2) -- (cgame);
\draw[arrow] (cgame) -- (cround1);
\draw[arrow] (cround1) -- (ccompleted);

\draw[arrow] (treat) -- (tgame);
\draw[arrow] (tgame) -- (tround1);
\draw[arrow] (tround1) -- (tcompleted);
\draw[arrow] (tcompleted) -- (t2);

\end{tikzpicture}%
}
\caption{Study flow and attrition.}
\label{fig:attrition}
\end{figure}

\subsection{Participants and Group Composition}
In this subsection we examine which respondents participated in the game and explore key group differences in socio-demographic variables and pre-existing attitudes. These analyses thus inform about the game's effectiveness to engage a wider audience, in accordance with the first part of the research question.

\begin{figure}[htb]
    \centering
    \includegraphics[width=\linewidth]{%
    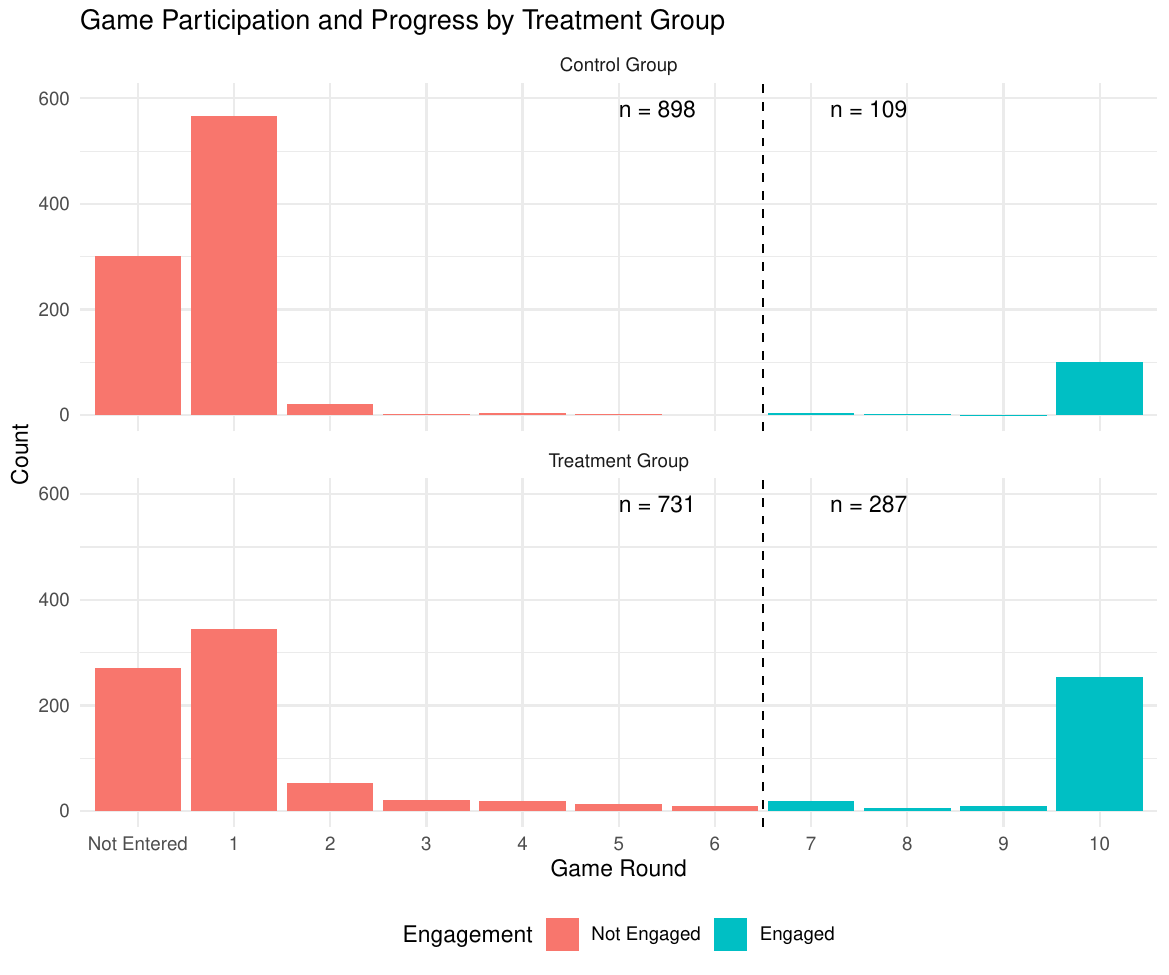}
    \caption{Rounds played in the game, by random treatment assignment.}
    \label{fig:game_prog}
\end{figure}

Figure \ref{fig:game_prog} shows the number of rounds played in the game by the treatment and control groups. While both groups exhibit a similar number of respondents entering the game, many more respondents in the control group exited within the first round, whereas a substantially larger share of the treatment group played until the final round. This indicates that participants who received an explicit invitation to play the game (i.e., the treatment group) exhibited greater engagement with it than those in the control group, who were only informed about the optional game after completing both parts of the survey.

To examine in more detail how engagement with the game relates to respondents’ socio-demographic and attitudinal characteristics, we define ``engagement'' as playing a minimum of seven (out of ten) rounds. This threshold is somewhat arbitrary but does not materially affect the results. As shown in Figure \ref{fig:game_prog}, most respondents either did not enter the game, exited before completing the first round, or completed all ten rounds.

\begin{figure}[htb]
    \centering
    \includegraphics[width=\linewidth]{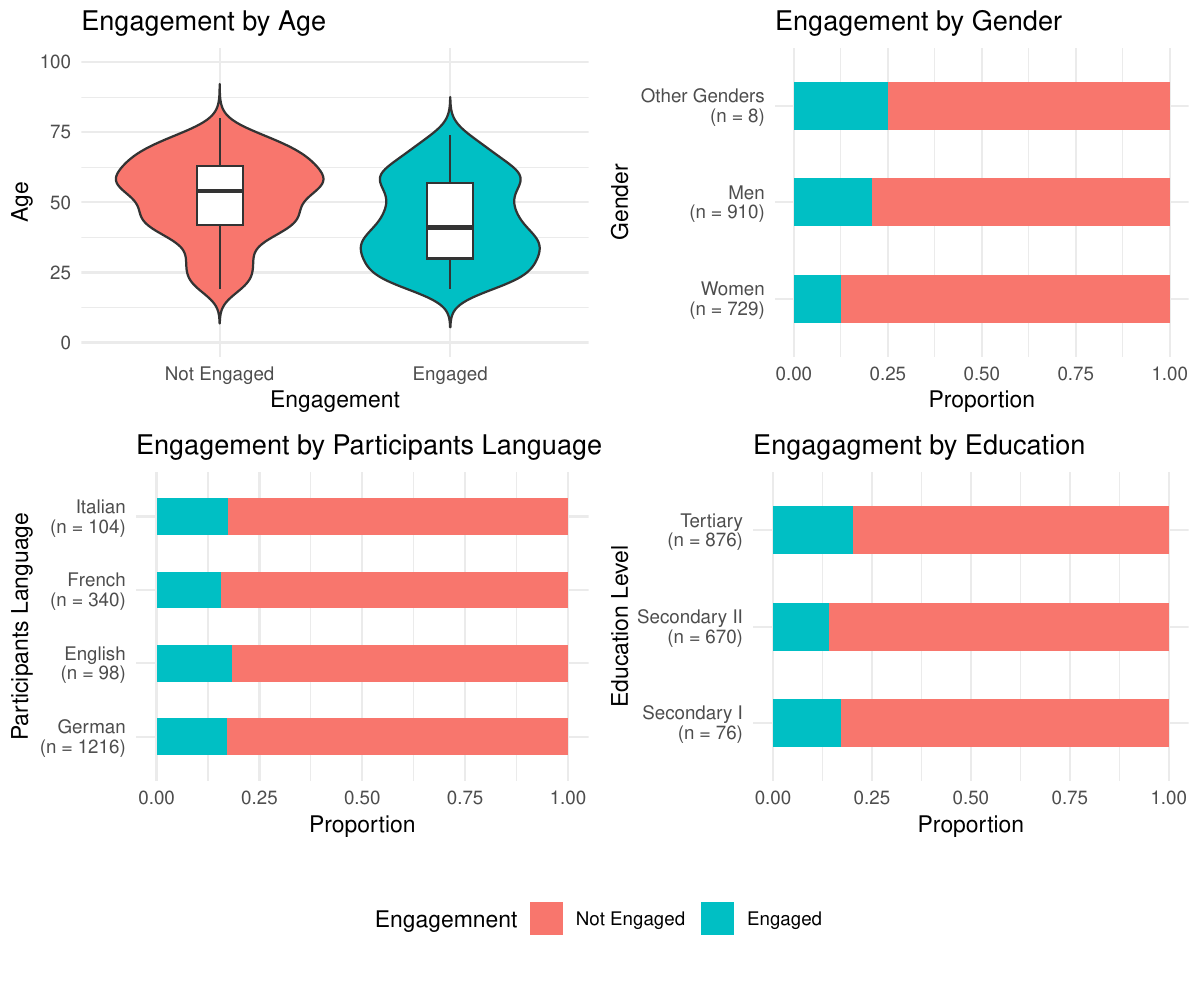}
    \caption{Game engagement by socio-demographic groups.}
    \label{fig:soz_dem_enga}
\end{figure}

To explore potential associations between game engagement patterns and socio-demographic variables, we consider age, gender, language, and educational attainment levels (see Figure \ref{fig:soz_dem_enga}). Looking at the age distribution of respondents by engagement, the median age for the group of engaged respondents has a lower median age (40 years old) compared to the non-engaged respondents (51 years old). This reflects similar patterns observed in the Canadian context by Yiannakoulias et al. \cite{yiannakoulias_decision_2020}, who employed a serious game in their study and noted a over-representation of respondents aged 20 to 39. Regarding gender, we see that respondents identifying as male or non-binary have slightly higher proportions of engagement with the game compared to female respondents. In contrast, there are no distinct patterns based on the language of the respondents. A slightly higher percentage of individuals from tertiary and secondary 1 education levels engaged with the game compared to those from secondary 2.

\begin{figure}[htb]
    \centering
    \includegraphics[width=\linewidth]{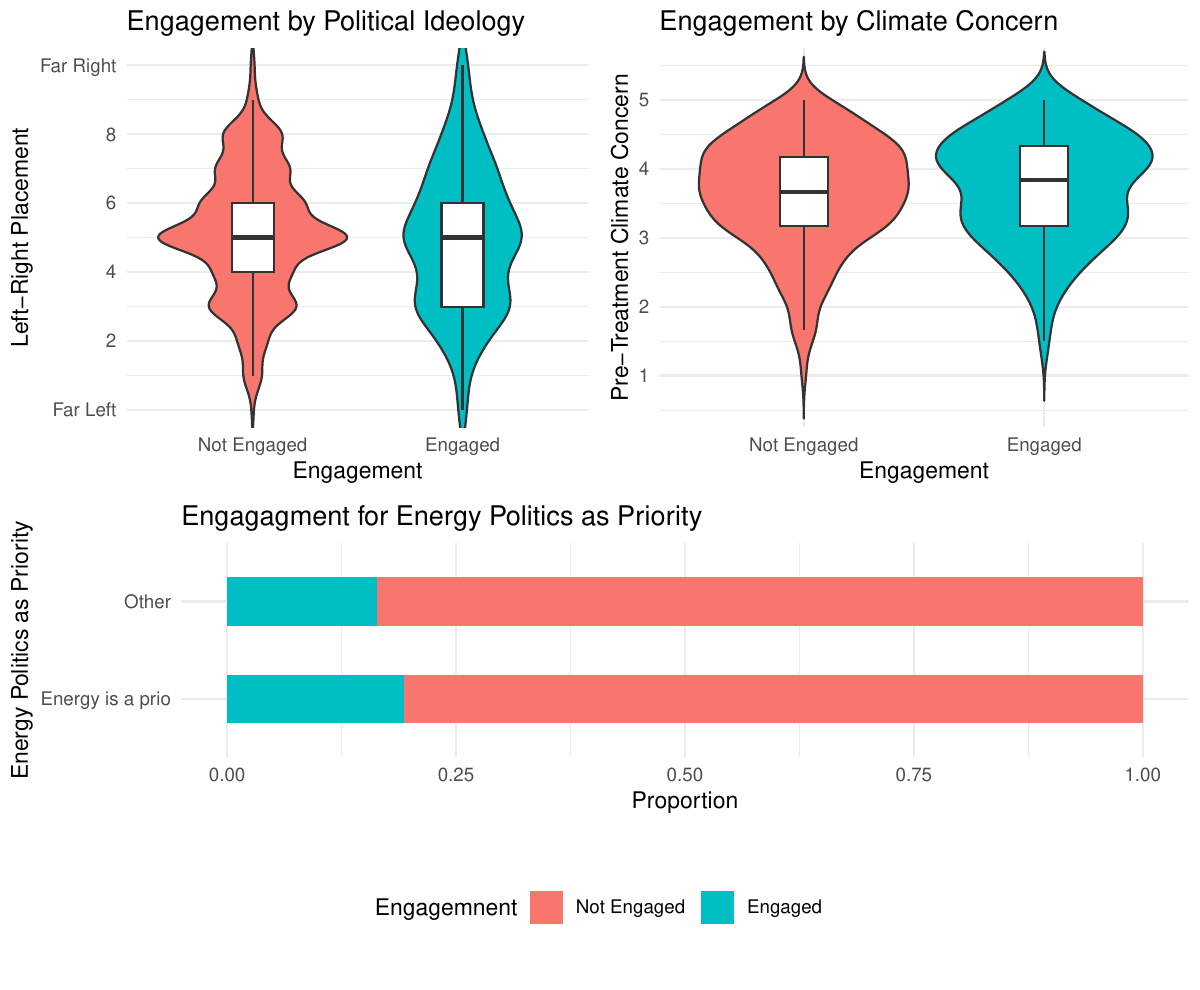}
    \caption{Game engagement by political attitudes.}
    \label{fig:poll_enga}
\end{figure}

Figure \ref{fig:poll_enga} depicts game engagement by varying political attitudes. In general, the findings reveal that the game engaged people with varying attitudes to a similar degree. However, it can be observed that participation was slightly higher among respondents with a politically left-leaning political orientation and among respondents with a higher climate concern, as well as among respondents who stated at the beginning of the survey that they consider energy politics a priority issue.

\subsection{Discussion}

Overall, the findings portray mixed conclusions. 
On one hand, assuming a ``glass half-empty'' position, it can be emphasised that an important share of respondents, despite being assigned to the treatment, did not engage with the game. Moreover, game engagement showed to be systematically dependent on socio-demographic and attitudinal factors. In view of the fact that especially less interested and less educated individuals were less likely to engage with the game, it can be concluded that the serious game was likely not able to reach and engage those individuals who would profit most from it. Rather, the capacity to form preferences on energy transition pathways and game engagement is likely driven by similar factors. This interpretation is supported by additional quantitative, pre-registered analyses on efficacy and policy support reported elsewhere \cite{Brückmann2026}. These findings indicate that there were no general treatment effects on either individual or collective efficacy beliefs. Accordingly, playing the game did not lead to a significant increase in people's perceptions of their own, or society's, ability to advance energy transitions \cite{Brückmann2026}.

On the other hand, however, a ``glass half-full'' position is also reasonable. Roughly a quarter of the treatment group did intensively engage with the game. Moreover, while the socio-demographic and attitudinal differences between players and non-players were statistically significant, they were in many cases not substantial in size. In particular, educational differences, which are known to be important drivers of higher sensitisation and acceptance of the energy transition and of receptivity to information information treatments, are small. These patterns suggest that the game was relatively successful in reaching individuals with lower educational resources.

\section{Conclusions}\label{sec:conc}

\subsection{Review}

In this paper, we motivated and described the development of a serious game around the Swiss energy transition and its integration into a large-scale population survey to evaluate the game's potential as an effective information tool. As the article documents, the game was successfully designed, developed and released as an interdisciplinary collaboration between political science, computational science, and professional game and graphic designs. This interdisciplinarity enabled us to create a game that is as realistic as possible in terms of the pathways for the Swiss energy system towards net zero. 

Integrating a serious game into a large-scale population survey proved to be a high-risk high-gain strategy. Previous research suggested that the immersive nature of a serious game could facilitate individual preference formation in the complex energy field, especially for those without strong prior knowledge or motivation to engage with the topic in a more traditional way. However, from a survey design perspective, it was unclear to what extent a broader population sample would be willing to engage with a serious game within a survey context. In general, this study demonstrated that the approach worked rather well, as only a relatively low number of respondents were deterred from the survey because of the game treatment. 

Based on the successful integration of the game into a population survey three important conclusions can be drawn. First, the game "works", i.e, it is in itself engaging and self-explanatory. User comments which were submitted alongside both the survey and game suggest that the game was evaluated positively by many (interested) players. Second, our analyses of socio-demographic and attitudinal characteristics of the players documents that the game was able to engage clearly beyond the stereotypical group of young males. While men were slightly more likely to engage with the game, the difference is not large. The same is true for education and age. It is also important to note that, even though the energy transition is strongly politicized in Switzerland, players with different ideological views engaged with the game in a similar way. Finally however, one caveat is that individuals less climate concern and those less interested in politics seem difficult to reach.

\subsection{Insights for Future Design}
\label{sec:insight1}

The question that guided this study was whether a serious game about energy transition can engage the broader population and can thus eventually be an effective tool for sensitisation and information. In addition to the empirical findings presented here, we want to return to this question from the perspective of game design. Our experience shows that the key element in design is to find the balance between the complexity of the information which is presented and the simplicity required for the game to be accessible and enjoyable for the players. While the interdisciplinary work was an important element for the game to become relevant, playable and feasible at the same time, the different disciplinary perspectives also are a challenge in terms of varying interests and expertise. 

Moreover, from the analysis of the results and from user feedback, it is clear that those interested and invested in the subject matter were happy to engage with the game. What is harder is to engage those who are not naturally drawn to the subject matter or intimidated by its complexity. 
We further note that the game was developed for a desktop-only browser environment. It became apparent during the survey that potential players were attempting to access the game using smartphone browsers, but were not able to continue due to the incompatibility of their devices. This represents a loss of engagement and data from a potentially valuable cohort to the study, most notably young people, who could, in turn, be hardest to engage through formal channels. %
We would advise future serious game designers 
Therefore, future studies should fully support smartphone users. The size and functionality of a touchscreen user interface thus places constraints on gameplay design which should be carefully considered.

\subsection{Research Implications}

A potential of the game that has not been exploited yet is the analysis of data collected in-game that records decisions relating to energy conversion
technology, nuclear power decommissioning and political policy will be further
analysed. For this, more players need to play the game to make this information a rich source for future analyses. Moreover, the model that provides the foundation
for the game can be used independently and will be used to generate data for further studies.

While our analyses pointed to the limitation in reaching the broader population, a fruitful approach for future dissemination and research activities could be the game's use in the educational sector. It can be assumed that students at the secondary II or tertiary level would be a suitable target group for the game. Further research could systematically evaluate whether the integration of the game in the classroom could be more effective than targeting the broader population.  

Together, these additional analyses and perspectives will provide further insight as to whether developing a serious game is worthwhile for social science research and knowledge transfer purposes.




\section*{Acknowledgments}
The research published in this report was carried out with the support of the 
Swiss Federal Office of Energy (SFOE) as part of the SWEET SURE consortium. 
The authors bear sole responsibility for the conclusions and the results presented in this publication. We thank all those who helped us testing the game and the participants recruited through the scientific survey.



 

\bibliographystyle{IEEEtran} 
\bibliography{bib1}

\end{document}